\documentclass[twocolumn,twocolappendix]{aastex631} 

\usepackage{amsmath}

\shorttitle{Silicate Grain Evolution in Massive Protostellar Disks}
\shortauthors{Yamamuro et al.}

\graphicspath{{./}{figures/}}

\begin{document}

\title{Massive Protostellar Disks as a Hot Laboratory of Silicate Grain Evolution}

\correspondingauthor{Ryota Yamamuro, Kei E. I. Tanaka}
\email{yamamuro.r.aa@m.titech.ac.jp, kei.tanaka@eps.sci.titech.ac.jp}

\author[0000-0001-7530-1359]{Ryota Yamamuro}
\affiliation{Department of Earth and Planetary Sciences, Tokyo Institute of Technology, Meguro, Tokyo, 152-8551, Japan}

\author[0000-0002-6907-0926]{Kei E. I. Tanaka}
\affiliation{Department of Earth and Planetary Sciences, Tokyo Institute of Technology, Meguro, Tokyo, 152-8551, Japan}
\affiliation{Center for Astrophysics and Space Astronomy, University of Colorado Boulder, Boulder, CO 80309, USA}
\affiliation{ALMA Project, National Astronomical Observatory of Japan, 2-21-1 Osawa, Mitaka, Tokyo 181-8588, Japan}

\author[0000-0002-1886-0880]{Satoshi Okuzumi}
\affiliation{Department of Earth and Planetary Sciences, Tokyo Institute of Technology, Meguro, Tokyo, 152-8551, Japan}

\begin{abstract}

Typical accretion disks around massive protostars are hot enough for water ice to sublimate.
We here propose to utilize the massive protostellar disks for investigating the collisional evolution of silicate grains with no ice mantle,
which is an essential process for the formation of rocky planetesimals in protoplanetary disks around lower-mass stars.
We for the first time develop a model of massive protostellar disks that includes the coagulation,  fragmentation, and radial drift of dust.
We show that
the maximum grain size in the disks is  limited by collisional fragmentation rather than by radial drift.
We derive analytic formulas that produce the radial distribution of the maximum grain size and dust surface density in the steady state.
Applying the analytic formulas to the massive protostellar disk of GGD27-MM1, where the grain size is constrained from a millimeter polarimetric observation, 
we infer that the silicate grains in this disk fragment at collision velocities above $\approx 10{\rm\:m\:s^{-1}}$.
The inferred fragmentation threshold velocity is lower than the maximum grain collision velocity in typical protoplanetary disks around low-mass stars, implying that coagulation alone may not lead to the formation of rocky planetesimals in those disks.
With future measurements of grain sizes in massive protostellar disks,
our model will provide more robust constraints on the sticking property of silicate grains.

\end{abstract}


\section{Introduction} \label{sec:intro}

Understanding the formation of kilometer-sized planetesimals from (sub)micron-sized grains is a critical issue in planet formation.
It is generally accepted that the grains grow into macroscopic aggregates through coagulation \citep{Johansen+2014,Drazkowska22}. However, whether the macroscopic aggregates can continue growing into kilometer-sized bodies by coagulation alone is much more uncertain because of their rapid radial drift  \citep{Weidenschilling+1977} and finite stickiness \citep{Blum+1993}.
If the grains are sticky enough, they may form fluffy aggregates and then grow into planetesimals more rapidly than they drift \citep{Okuzumi+2012,Kataoka+13}.
Otherwise, other mechanisms that rapidly concentrate dust, such as the streaming and gravitational instabilities \citep[e.g.,][]{Goldreich+1973,Youdin+2005,Johansen2007,Youdin+2011}, are necessary for planetesimal formation.
Therefore, in order to understand how planetesimals form, one must know how sticky the grains are.

A major concern in planetesimal formation studies is that the stickiness of silicate grains is highly uncertain. Silicates are the ultimate building blocks of rocky bodies, and therefore knowing their stickiness is necessary for understanding how rocky planets like the Earth form. Early grain sticking models \citep{Chokshi+93,Dominik+97} predicted that micron-sized rocky grains can stick at collision velocities up to $\sim 0.1~{\rm m~s^{-1}}$. However, laboratory experiments typically show that micron-sized silica grains can stick even at  $1{\rm\: m\:s^{-1}}$ \citep[][]{Poppe+2000,Blum+00}. The source of the large discrepancy is yet to be confirmed, but recent studies \citep{Kimura+15,Steinpilz+19} suggest that the early models underestimated the surface energy (a measure of the strength of the intermolecular forces) of silicates. The stickiness of dust aggregates also depends on the size of the constituent grains \citep{Chokshi+93,Dominik+97}. For instance, aggregates made of $0.1{\rm\:\micron}$-sized interstellar silicate grains may stick at velocities up to $50\:\rm m\:s^{-1}$ \citep{Kimura+15}.
Because of this large uncertainty, one cannot yet conclude whether coagulation of silicate aggregates into kilometer-sized bodies is possible or not in protoplanetary disks, where the maximum grain collisional velocity is $\approx 20$--$50{\rm\:m\:s^{-1}}$ unless strong turbulence is present \citep[e.g., Figure 4 of][]{Johansen+2014}.

Given this importance, it is desirable to constrain the stickiness of silicate grains from observations of protoplanetary disks. 
However, it is challenging to do this because the disks around low-mass stars ($\sim1\:M_\odot$, $1\:L_\odot$) are cold except in the innermost region or in the accretion outburst phase \citep[e.g.,][]{Liu2021}. 
In these disks, the radius of the snowline, where water ice sublimates, is only a few au or even smaller  \citep[e.g.,][]{Mori+21},
and it is difficult to spatially resolve their silicate-dust regions inside the snowlines.
On the other hand, the snowline radii around massive protostars ($\ga10\:M_\odot$, $\ga10^3\:L_\odot$) are typically much large as several hundred au or more due to the intense stellar radiation and high accretion heating.
Thus, silicate-dust regions in massive protostellar disks can be resolved by current telescopes, such as Atacama Large Millimeter/submillimeter Array (ALMA),
although massive star-forming regions are typically distant as several kpc.
In particular, the 1.14-mm polarimetric ALMA observations by \cite{Girart+18} found that the maximum grain size is as large as several hundred microns in the massive protostellar disk of GGD27-MM1.
Given the high luminosity and the accretion rate of GGD27-MM1 \citep[$\sim10^4\:L_\odot$ and $10^{-4}\:M_\odot{\rm\:yr^{-1}}$,][]{Anez+2020},
its water snowline is likely to present outside of the accretion disk,
suggesting that the observed large grain size is an excellent indicator of the coagulation of silicate dust.

Motivated by the detection of large-size silicate grains in GGD27-MM1,
we develop the first theoretical model of dust coagulation in massive protostellar disks in this paper.
Massive protostellar disks have not gotten significant attention as the site of dust coagulation, maybe because Earth-like habitable planets would not form around massive stars due to their intense radiation field and short lifetimes.
However, massive protostellar disks are excellent observational targets to investigate the coagulation of silicate grains.
Furthermore, ALMA observations have started to discover accretion disks around massive protostars in recent years \citep[e.g.,][]{Ginsburg+18, Maud+19, Motogi+19, Tanaka+2020, Johnston+2020}.
We need to know the appropriate dust size (i.e., its opacity) to unravel their physical properties from observations, especially dust mass.
 Massive protostars have orders of magnitude higher luminosities ($\sim10^3L_\odot$) and accretion rates ($\sim10^{-3}$--$10^{-4}{M}_\odot\:\mathrm{yr}^{-1}$) than low-mass protostars \citep{Hirota+18}.
Thus, dust coagulation in massive protostellar disks is nontrivial:
how much dust grains can grow, what process limits their coagulation, and how dust coagulation impact disk physical conditions.
A theoretical model is required to understand observation results and constrain the sticking property of silicate grains.

The structure of this paper is as follows.
We describe our disk model with dust coagulation in Section \ref{sec:Method}.
We derive the analytic formula of the size limit of dust grains in Section \ref{sec:Analytic},
and show the results of numerical calculations in Section\:\ref{sec:Result}.
We provide the discussion in \ref{sec:discc},
where we constrain the threshold fragmentation velocity of silicate dust by comparing our model with the observed grain size of GGD27-MM1.
In Section\:\ref{sec:sum}, we present the summary of this paper.

\section{Model} \label{sec:Method}
\begin{figure*}
\gridline{\fig{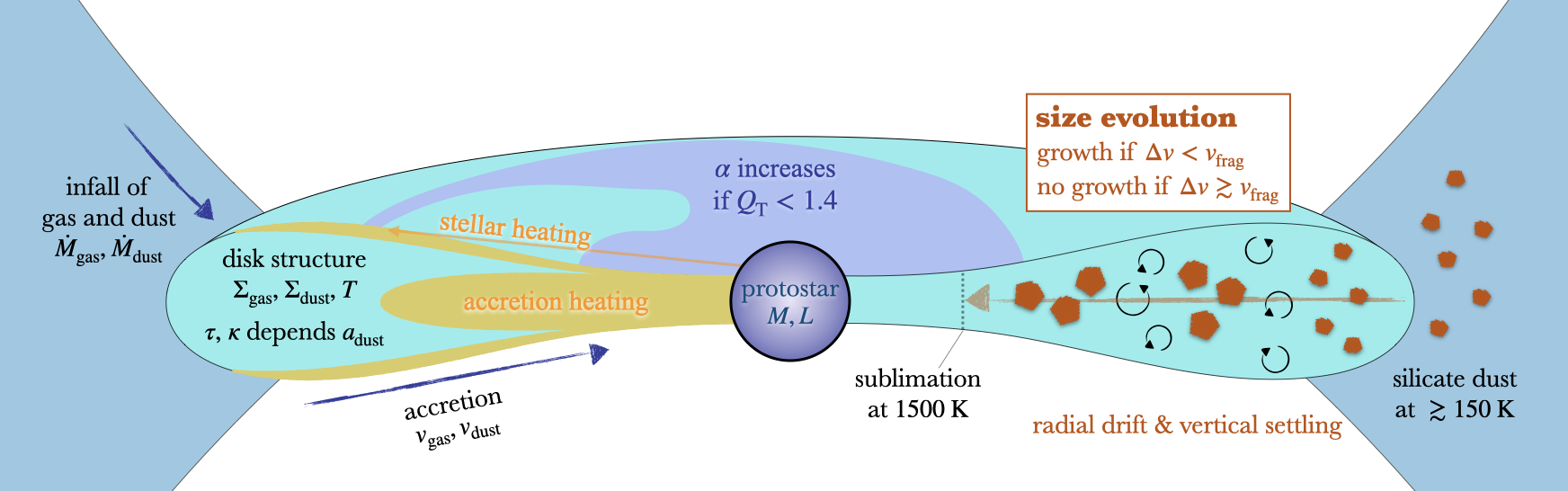}{1\textwidth}{}}
\caption{Schematic illustration showing the overview of our disk model with dust evolution.
{\it (left)} The disk is in a steady state with a supply of gas and dust from the envelope with the rates of $\dot{M}_{\rm gas}$ and $\dot{M}_{\rm dust}$.
The heating from the stellar irradiation and the accretion heating are considered.
For the accretion heating, we take into account the vertical optical depth $\tau$, which depends on the dust coagulation.
We use the $\alpha$ viscosity to treat the angular momentum transfer of the gas disk.
The $\alpha$ value increases where the disk is marginally unstable with the Toomre parameter of $Q<1.4$.
{\it (right)} For dust evolution, we consider dust collisional growth and fragmentation, as well as radial drift and vertical settling.
The dust coagulation is suppressed if the collisional velocity is as high as or higher than the fragmentation velocity, i.e., $\Delta v\ga v_{\rm frag}$.
The entire disk is the silicate-dust region inside the water snowline.
The numerical calculation is executed from the disk outer radius $r_{\rm d}$ to the inner radius of $10{\rm\:au}$, or the dust sublimation radius with $1500{\rm\:K}$.
\label{fig:overview}}
\end{figure*}

We present our model for massive protostellar disks with dust coagulation,
which is schematically shown in Figure \ref{fig:overview}.
We consider an accretion disk with a supply of gas and dust from the surrounding envelope.
The disk's interior temperature is calculated by considering both stellar irradiation and accretion heating.
For dust grains, we take into account collisional growth and fragmentation as well as radial drift and vertical settling.
We detail each process in the following subsections. 

\subsection{Gas Disk} \label{sec:gas}
We assume that the mass flux from the envelope to the disk is approximately constant in time and axisymmetric. We then decompose the disk structure into a steady, axisymmetric part and time-dependent, non-axisymmetric disturbances. The latter may include density waves produced by the disk's self-gravity and turbulence. Below, we focus on the former structure and model the disk as being steady and radially one-dimensional.
However, we also account for (time-averaged)
angular momentum transport induced by the disk's self-gravity.

The disk's rotation is assumed to be nearly Keplerian. In fact, the gas rotation is slightly slower than Keplerian owing to the outward pressure support. This small deviation is usually negligible regarding the gas disk structure.
However, the sub-Keplerian motion must be taken into account when computing the radial motion of the grains  (see Section\:\ref{sec:dust}).

The equation of continuity for an axisymmetric gas disk reads \begin{equation}
\label{gasrenzoku}
\frac{\partial \Sigma_\mathrm{gas}}{\partial t}+ \frac{1}{r}\frac{\partial}{\partial r}(rv_{r,\mathrm{gas}}\Sigma_\mathrm{gas})=0,
\end{equation}
where $\Sigma_\mathrm{gas}$ and $v_{r,\mathrm{gas}}$ are the gas surface density and velocity, and $r$ is the distance from the central protostar.
Assuming steady accretion, we obtain 
\begin{equation}
\label{gas1}    
\dot{M}_\mathrm{gas} \equiv 2\pi r|v_{r,\mathrm{gas}}|\Sigma_\mathrm{gas}=\mathrm{const.},
\end{equation}
where $\dot{M}_\mathrm{gas}$ is the gas accretion rate.

We use the $\alpha$ viscosity to describe the gas angular momentum transport within the disk \citep{Shakura+1973}. 
The accretion velocity in the steady viscous disk can then be written as
\begin{equation}\label{eq:vgr}
v_{r\mathrm{,gas}} =-\frac{3\alpha c_{\rm s}^2}{2v_\mathrm{K}},
\end{equation}
where $\alpha$ is the dimensionless viscosity parameter and
\begin{align}
c_\mathrm{s}=\sqrt{\frac{k_\mathrm{B}T}{m_\mathrm{gas}}},\\
v_\mathrm{K}=\sqrt{\frac{GM}{r}},
\end{align}
are the sound speed and Keplerian velocity, respectively, with $M$, $G$, $k_\mathrm{B}$, and $m_\mathrm{gas}$ being the stellar mass, gravitational constant, Boltzmann constant, and mean molecular mass.
The disk's vertically averaged temperature $T$ is approximated by the midplane temperature.
We use $m_\mathrm{gas}=3.9\times10^{-24}\:\mathrm{g}$, assuming a mean molecular weight of $2.34$.

A massive protostellar disk can be gravitationally unstable.
In such a disk, the gravitational instability (GI) induces spiral density waves, whose torque transfer the disk's angular momentum.
We therefore decompose the viscosity parameter $\alpha$ into two components:
\begin{equation} \label{hanpuku4}
\alpha=\alpha_\mathrm{GI}+\alpha_\mathrm{floor},
\end{equation}
where $\alpha_\mathrm{GI}$ and $\alpha_\mathrm{floor}$ represent the contributions from the GI and any other angular-momentum transport mechanisms, respectively.
Following \citet{Zhu+10}, we model the former as 
\begin{equation}\label{aGI}
\alpha_\mathrm{GI}=\exp\left(-Q^4\right),
\end{equation}
where
\begin{equation}\label{ToomreQ}
Q=\frac{c_\mathrm{s}\Omega_{\mathrm K}}{\pi G \Sigma_\mathrm{gas}},
\end{equation}
is the Toomre parameter for the GI \citep{Toomre1964} with $\Omega_{\rm K} = v_{\rm K}/r$ being the Keplerian angular velocity.
Equation~\eqref{aGI} mimics the efficient angular momentum transport by spiral density waves that develop when the disk is marginally unstable state of $Q\la 1.4$ \citep[e.g.,][]{Boley+06,Zhu+10}.
Even in gravitationally stable disks ($Q>1.4$), angular momentum transport by hydrodynamical and magnetohydrodynamical turbulence and by large-scale magnetic fields can still occur (see \cite{Lesur2022} for a recent review of potential angular momentum transport mechanisms in protoplanetary disks).
Following \citet{Tanaka+2014}, we take $\alpha_{\rm floor}=0.01$.

To determine the disk temperature $T$, we assume radiative equilibrium and consider heating by stellar irradiation and disk accretion.
We approximate $T$ as 
\begin{equation} \label{hanpuku5}
T=\left(T_\mathrm{irr}^4+T_\mathrm{acc}^4\right)^{1/4},
\end{equation}
where $T_\mathrm{irr}$ and $T_\mathrm{acc}$ are the temperatures in the limits where irradiation heating and accretion heating dominate, respectively. 
The former can be estimated as \citep{Chiang1974,Kusaka}
\begin{equation}\label{Tirr}
T_\mathrm{irr}=1.5\times10^2\left(\frac{L}{L_\odot}\right)^{2/7}
\left(\frac{M}{M_\odot}\right)^{-1/7}
\left(\frac{r}{\mathrm{au}}\right)^{-3/7} \:\mathrm{K},
\end{equation}
where $L$ is the stellar luminosity.
Equation~\eqref{Tirr} assumes that the disk is optically thick to the starlight, which is valid unless small dust grains are heavily depleted.
With vertically uniform viscosity,
the accretion-dominated temperature is \citep{Hubeny+1990}
\begin{equation}
\label{Tacc}
T_\mathrm{acc}=\left(\frac{\sqrt{3}}{4}+\frac{3}{16}\tau\right)^{1/4}\left(\frac{3GM\dot{M}_\mathrm{gas}}{8\pi\sigma_\mathrm{SB}r^3}\right)^{1/4},
\end{equation}
where $\sigma_\mathrm{SB}$ is the Stefan--Boltzmann constant, and $\tau/2$ is the  disk's vertical optical depth to its own thermal emission, measured from infinity to the midplane.
An optically thicker disk is hotter because of inefficient cooling.

\subsection{Optical Depth}\label{subsec:op}
Dust grains are the dominant source of the disk opacity. 
For grains with size distribution, the disk's vertical optical depth can be written as
\begin{equation}\label{eq:tau}
\tau(r)=\int_{a_\mathrm{min}}^{a_\mathrm{max}} \sigma_\mathrm{abs}(a)\frac{dN_{\rm dust}(a, r)}{da}da,
\end{equation}
where $\sigma_\mathrm{abs}(a)$ is the monochromatic absorption cross-section of a grain with the radius $a$, $dN_{\rm dust}(a,r)/da$ is the number surface density of grains per unit grain radius, and $a_{\rm min}$ and $a_{\rm max}$ are the minimum and maximum grain radii, respectively. 
The grain size distribution is assumed to follow a power law
\begin{equation}\label{eq:distri}
\frac{dN_{\rm dust}}{da}=Ca^{-3.5}.
\end{equation}
Here, the normalization factor $C$ satisfies 
\begin{equation}
\Sigma_{\rm dust}
=\int_{a_\mathrm{min}}^{a_\mathrm{max}} \frac{4\pi a^3}{3}\rho_\mathrm{int}\frac{dN_{\rm dust}}{da}da,
\end{equation}
where $\Sigma_{\rm dust}$ is the dust surface density, and $\rho_\mathrm{int}=3.0{\rm\:g\:cm^{-3}}$ is the internal density of silicate grains.
Equation~\eqref{eq:distri} well approximates the high-mass end of the grain size distribution in  coagulation--fragmentation equilibrium in turbulent disks \citep{Birnstiel+2012}. 
We take $a_{\rm min} = 0.1\:\micron$ because grains smaller than this size generally grow quickly thanks to Brownian motion \citep{Birnstiel+11}. The maximum grain size $a_{\rm max}$ is determine from a dust coagulation model (Section \ref{sec:amax}).

For $\sigma_{\rm abs}$, we use a crude approximation 
\citep[e.g.,][]{Ivezic97,Ormel14,Fukuhara21},
\begin{equation}\label{eq:kappa}
\sigma_\mathrm{abs}(a)=\pi a^2\:\mathrm{min}
\left(1,\frac{2\pi a}{\lambda_\mathrm{peak}}\right),
\end{equation}
where $\lambda_\mathrm{peak}=10(300{\rm\:K}/T)\:\micron$ is the peak wavelength of the Planck function with $T$.
As long as $T \ga 100\rm\:K$, Equation~\eqref{eq:kappa} approximates the real grain opacity to within a factor of a few \citep{Dullemond22}. 
For convenience in presenting the results, we introduce the absorption opacity per gas mass as $\kappa_\mathrm{abs}=\tau/\Sigma_\mathrm{gas}$.

\subsection{Dust Surface Density}\label{sec:dust}
We employ the single-size approach of \cite{Sato+2016} to compute the dust evolution in the disk.
This approach assumes that grains of one characteristic size dominate the dust mass surface density at each orbit $r$. 
For the grain size distribution given by Equation~\eqref{eq:distri}, the mass-dominating grains are those with $a \sim a_{\rm max}$\footnote{\label{footnote1}Formally, the size of the mass-dominating grains can be calculated as (Equation (A.5) of \citealt{Sato+2016}; see also Ormel \& Spaans 2008) 
\begin{equation}
a_{\rm p} = \left(\frac{\int_{a_{\rm min}}^{a_{\rm max}}a^6 (dN_{\rm dust}/da)da}{\int_{a_{\rm min}}^{a_{\rm max}}a^3 (dN_{\rm dust}/da)da}\right)^{1/3}.
\end{equation}
For $dN_{\rm dust}/da$ given by Equation\:\eqref{eq:distri} with $a_{\rm max} \gg a_\mathrm{min}$, one has $a_\mathrm{p} \approx 0.52a_\mathrm{max}$.}.
In our model, we identify the mass-dominating grains with the largest grains.

In the single-size approach, the dust surface density $\Sigma_{\rm dust}$ in an axisymmetric disk generally obeys
\begin{equation}
 \label{dustrenzoku}
 \frac{\partial \Sigma_{\rm dust}}{\partial t}+ \frac{1}{r}\frac{\partial}{\partial r} (rv_{r,\mathrm{dust}}\Sigma_{\rm dust})=0  ,
\end{equation}
where $v_{r,\mathrm{dust}}$ is the radial velocity of the largest grains.
Because the radial flow of the dust from the envelope to the star is assumed to be steady, Equation~\eqref{dustrenzoku} yields  
\begin{equation}
\label{dust1}    
\dot{M}_\mathrm{dust}= 2\pi r|v_{r,\mathrm{dust}}|\Sigma_{\rm dust}=\mathrm{const.} ,
\end{equation}
where $\dot{M}_\mathrm{dust}$ is the dust accretion rate.
We take $\dot{M}_\mathrm{dust}$ to be $\dot{M}_\mathrm{gas}$ times the interstellar dust-to-gas mass ratio of $f_{\rm dg, ISM} = 0.01$.
We note that the ratio of $\dot{M}_\mathrm{gas}$ and $\dot{M}_\mathrm{dust}$ is constant in the entire disk (Equations \ref{gas1} and \ref{dust1}),
but the ratio of $\Sigma_\mathrm{gas}$ and $\Sigma_\mathrm{dust}$ is not, because the radial velocities of gas and dust are different due to the dust radial drift.

The dust radial velocity is determined by gas drag in the disk. We use the expression \citep{Takeuchi+2002}
\begin{equation}
v_{r,\mathrm{dust}}=\frac{v_{r,\mathrm{gas}}} {1+{\mathrm{St}}^2}     -    \frac{2\mathrm{St}} {1+{\mathrm{St}}^2}\eta v_\mathrm{K},
\label{v_r_dust}
\end{equation}
where ${\rm St}$ is the largest grains' stopping time (see below) normalized by the inverse Keplerian frequency, i.e., ${\rm St} = \Omega_{\rm K}t_{\rm s}$, and 
\begin{equation}
\eta \equiv -\frac{1}{2}\left(\frac{c_{\rm s}}{v_\mathrm{K}}\right)^2
\frac{d\mathrm{ln}P}{d\mathrm{ln}r}
\end{equation}
is a dimensionless parameter characterizing the disk's sub-Keplerian motion \citep{Adachi+1976}, with $P$ being the gas pressure.
Note that $\eta$ is positive in a smooth gas disk with radially decreasing pressure. 
The parameter ${\rm St}$, called the Stokes number, measures how strongly the grains' motion is coupled to the gas motion, with the limit ${\rm St} \to 0$ corresponding to the extreme case where the grains and gas move at the same velocity. 
On the right-hand side of Equation\:(\ref{v_r_dust}), the first term simply represents the radial motion induced by gas accretion. The second term represents the so-called radial inward drift in a sub-Keplerian rotating disk: grains feel the headwind of the background gas, lose angular momentum, and fall toward the central star \citep{Whipple1972,Adachi+1976,Weidenschilling+1977}. For simplicity, we fix $d\mathrm{ln}P/d\mathrm{ln}r=-24/7$, which exactly applies when $T = T_{\rm irr}$ and $Q = \rm{const}$. \citep{Tsukamoto+17}.

The stopping time characterizes the timescale on which the grains' velocity relaxes into the terminal velocity under gas drag. 
If the largest grains are smaller than the gas molecules' mean free path, they obey Epstein's law
\begin{equation}
t_\mathrm{s}=
\frac{\rho_\mathrm{int}a_{\rm max}}{\rho_\mathrm{gas}v_\mathrm{th}},
\end{equation}
where $\rho_\mathrm{gas}$ is the gas density, and 
$v_\mathrm{th}=\sqrt{8/\pi}c_{\rm s}$ is the molecules' mean thermal speed.
Following \citet{Birnstiel+2010}, we approximate $\rho_{\rm gas}$ with the midplane density in an isothermal disk,
$\rho_{\rm gas} = \Sigma_{\rm gas}/(\sqrt{2\pi}H_{\rm gas})$, where $H_{\rm gas} = c_{\rm s}/\Omega_{\rm K}$ is the gas scale height. We then have
\begin{equation}\label{eq:St}
\mathrm{St}=\frac{\pi \rho_\mathrm{int}a_{\rm max}}{2\mathrm{\Sigma_\mathrm{gas}}}.
\end{equation}
In our calculations, most of the largest grains are smaller than the molecules' mean free path of $\lambda_\mathrm{mfp}={m_\mathrm{gas}}/(\sigma_\mathrm{mol}\rho_\mathrm{gas})$, 
where $\sigma_\mathrm{mol}=2.0\times10^{-15}\:\mathrm{cm^{-2}}$ is the collision cross-section of the gas molecules.
Therefore, the use of Epstein's law is valid.

\subsection{Maximum Grain Size}\label{sec:amax}
We take into account collisional fragmentation depending on the grains' collision velocity. 
The collisional growth/fragmentation and radial drift control the maximum grain size at each disk radius.

In the single-size approach, the mass of the largest grains, $m_{\rm max}=(4\pi/3) \rho_\mathrm{int} a_{\rm max}^3$, obeys \citep{Sato+2016,Okuzumi16}
\begin{equation}
\label{dustgrow}
\frac{\partial m_{\rm max}}{\partial t}+v_{r,\mathrm{dust}}\frac{\partial m_{\rm max}}{\partial r} 
 =  \xi_\mathrm{frag} \frac{m_{\rm max}}{t_\mathrm{coll}},
\end{equation}
where $t_\mathrm{coll}$ is the largest grains' mean collision time, and $\xi _{\rm frag}$ is the fractional change of $m_\mathrm{\rm max}$ upon a single collision, accounting for collisional fragmentation.
Because we consider steady state, Equation (\ref{dustgrow}) reduces to an ordinary differential equation for $a_{\rm max}$,
\begin{equation}
\frac{da_{\rm max}}{dr}
=\xi_\mathrm{frag} \frac{a_{\rm max}}{3t_\mathrm{coll} v_\mathrm{r,dust}}.\label{eq:da}
\end{equation}

Following \citet{Okuzumi+Hirose+2011} and \citet{Okuzumi16}, we model $\xi _{\rm frag}$ as
\begin{equation} \label{xi_frag}
\xi_\mathrm{frag}=
    \mathrm{min} \left(1,-\frac{\mathrm{ln}(\Delta v/v_\mathrm{frag})}{\mathrm{ln}5}\right) ,
\end{equation}
where $\Delta v$ is the mean collision velocity and $v_{\rm frag}$ is the threshold velocity above which a net mass loss of the largest grains occurs.
Hereafter, we simply call $v_{\rm frag}$ the fragmentation velocity. 
Equation~\eqref{xi_frag} is based on dust aggregate collision simulations by \citet{Wada+09}, which show that $\xi_{\rm frag}$ is $\approx 1$ at $\Delta v \la 0.2v_{\rm frag}$ and decreases approximately logarithmically with $\Delta v$ (see Figure 11 of \citealt{Wada+09}).

The collision time $t_\mathrm{coll}$ is the time taken for dust grains of the same size to collide.
With the number density of dust grains of $n_\mathrm{dust}$, the collision time is given as
\begin{equation}
t_\mathrm{coll}=\frac{1}{4\pi a_{\rm max}^2 n_\mathrm{dust} \Delta v}.
\end{equation}
where
\begin{equation}
n_\mathrm{dust}=\frac{\Sigma_{\rm dust}}{\sqrt{2\pi}H_\mathrm{dust}m_{\rm max}},
    \end{equation}
is the number density of dust grains, and
\begin{equation}\label{eq:Hdust}
H_\mathrm{dust}=\left(1+ \frac{ \mathrm{St} }{\alpha} \right)^{-1/2} H_\mathrm{gas},
\end{equation}
is the dust scale-height
determined by the balance between vertical settling and turbulent diffusion \citep{Dubrulle95}. 
In turbulent disks as considered in this study, turbulence is the main source of relative velocity \citep{Johansen+2014}. 
Based on the analytic study by \citet{Ormel+2007}, we model the turbulence-induced relative velocity as 
\begin{subequations}
\label{deltav}
\begin{align}
\Delta v &=\min \left( \Delta v_1, \Delta v_2 \right), \label{deltav0}\\
&\Delta v_1 = \frac{1}{2} \alpha^{1/2} c_\mathrm{s} \mathrm{St}{\rm\:Re}^{1/4},
\label{deltav1}
\\
&\Delta v_2 = \sqrt{2.3\alpha c_{\rm s}^2 \mathrm{St}},
\label{deltav2}
\end{align}
\end{subequations}
where ${\rm Re}=\alpha \Sigma_\mathrm{gas}\sigma_\mathrm{mol}/(2m_\mathrm{gas}$) is the turbulent Reynolds number.
Equation~\eqref{deltav} continuously connects the asymptotic expressions for $\Delta v$ in the limits of ${\rm St}\ll{\rm Re}^{-1/2}$ and $\gg{\rm Re}^{-1/2}$, i.e., $\Delta v_1$ and $\Delta v_2$
\citep[see Equations (27) and (28) of][]{Ormel+2007}.
The factors $1/2$ and $2.3$ in Equations~\eqref{deltav1} and \eqref{deltav2} assume
that the largest grains collide mainly with smaller grains of Stokes number ${\rm St}' \approx (1/2)\mathrm{St}$ \citep{Okuzumi16}.

It should be noted that we employ a single $\alpha$ value to describe both the angular momentum transfer efficiency (the so-called viscosity) and the velocity dispersion of gas turbulence. While the former drives gas accretion (Equation~\eqref{eq:vgr}), the latter induces dust diffusion (Equation~\eqref{eq:Hdust}) and collision (Equation~\eqref{deltav}). 
The two physically distinct quantities would be comparable in magnitude if nearly isotropic turbulence is the main driver of the accretion. However, they can have significantly different values in self-gravitating disks, where the accretion stress from the self-gravitating spirals dominates over the Reynolds stress from turbulence \citep{Baehr+21B}. Observations of protoplanetary disks that exhibit a high level of dust settling also point to dust diffusivity much lower than the angular momentum transport efficiency \citep{Pinte2016,Ribas+2020,Villenave2022ApJ}.
In Section \ref{sec:caveats},  we assess the effects of $\alpha_{\rm acc}\neq\alpha_{\rm turb}$ on our results.

\subsection{Boundary Conditions and Parameter Choices}\label{sec:basePara}
We numerical solve Equation~\eqref{eq:da} from the disk outer edge at $r=r_{\rm d}$ inward. The maximum grain size at $r=r_{\rm d}$ is taken to be $a_{\rm max}=a_{\rm min}=0.1\:\micron$.

We consider a wide range of physical parameters (Table \ref{tab:model}).
In the fiducial model, we consider the central massive protostar of $M=20\:M_\odot$ and $L=4\times 10^4\:L_\odot$ with the disk of $r_{\rm dist}=150{\rm\:au}$.
For the gas accretion rate, we explore over two orders of magnitude $\dot{M}_\mathrm{gas}=10^{-5}$--$10^{-3}\:{M_\odot\:\mathrm{yr}^{-1}}$,
selecting $10^{-4}\:{M_\odot\:\mathrm{yr}^{-1}}$ as the fiducial value.
As explained in Section \ref{sec:intro}, there is a large degree of uncertainty in the fragmentation velocity of silicate grains.
To investigate how the $v_{\rm frag}$ value affects dust coagulation, we vary $v_{\rm frag}$ between $1$--$100{\rm\:m\:s^{-1}}$, with $10{\rm\:m\:s^{-1}}$ being the fiducial value.

Our ultimate goal is to constrain the sticking property of silicate grains by comparing our model and observations.
To date, the massive protostellar disk of GGD27-MM1 is the only object where the maximum silicate grain size is observationally constrained \citep{Girart+18}.
Thus, we also perform model calculations with the parameters of GGD27-MM1 \citep{Anez+2020}, searching for the value of $v_{\rm frag}$ that yields the observationally inferred grain size, analytically in Section \ref{giron:1} and numerically in Appendix \ref{sec:appA}.

\begin{deluxetable*}{lccccc}
\tablecaption{Model Parameters Investigated by Numerical Calculations\label{tab:model}}
\tablehead{
\colhead{Model}&\colhead{Stellar Mass}&\colhead{Stellar Luminosity}&\colhead{Gas Accretion Rate} &\colhead{Disk Radius} & \colhead{Fragmentation Velocity}\\
\colhead{}&\colhead{$M$\:($M_\odot$)}&\colhead{$L$\:($L_\odot$)}&\colhead{$\dot{M}_{\rm gas}$\:($M_\odot\:\mathrm{yr}^{-1}$)} &\colhead{$r_{\rm d}$\:(au)} & \colhead{$v_{\rm frag}$\:($\rm m\:s^{-1}$)}
}
\startdata
Fiducial model in Section \ref{subRe:fd} & 20 & $4\times10^4$ & $1\times10^{-4}$ & 150 & 10 \\
Comparison models in Section \ref{subRe:vfrag} & 20 & $4\times10^4$ & $1\times10^{-4}$ & 150 & 1,~100 \\
Comparison models in Section \ref{subRe:Mdot} & 20 & $4\times10^4$ & $10^{-5},~10^{-3}$ & 150 & 10 \\
\hline
GGD27-MM1 models in Appendix \ref{sec:appA}$^a$ & 20 & $1.4\times10^4$ & $7\times10^{-5}$ & 168 & 1,\:10,\:100 \\
\enddata
\tablecomments{We fix the values of the grain density and the interstellar dust-to-gas mass ratio as $\rho_{\rm int}=3{\rm\:g\:cm^{-3}}$ and $f_{\rm dg,ISM}=0.01$ for all models.
$^a$The parameters of the GGD27-MM1 models come from \citet{Anez+2020}, except for the fragmentation velocity.}
\end{deluxetable*}

\section{Analytic Estimates} \label{sec:Analytic}

As already demonstrated by previous studies \citep[e.g.,][]{Birnstiel09,Birnstiel+2012,Okuzumi16}, one can easily estimate the maximum size of dust grains when their growth is limited by either collisional fragmentation or radial drift. Our numerical calculations presented in Section~\ref{sec:Result} show that  fragmentation is the dominant growth barrier (see Section~\ref{sec:Result}) in massive protostellar disks. 
Before presenting the numerical results, we here derive analytic expressions for the fragmentation-limited grain size in self-gravitating disks, making use of the fact that the gas density and temperature in the disks can be expressed in terms of the Toomre parameter $Q$. Such expressions are useful for 
understanding how the maximum grain size in massive protostellar disks depends on various parameters.

We begin by using Equations (\ref{gas1}) and (\ref{ToomreQ}) to rewrite the gas surface density and sound velocity as
\begin{align}
\Sigma_\mathrm{gas}=\frac{\dot{M}_\mathrm{gas}\Omega_{\rm K}}{3\pi \alpha c_\mathrm{s}^2}
&=\left( \frac{\dot{M}_\mathrm{gas} \Omega_\mathrm{K}^3}{3 \pi^3 G^{2}\alpha Q^{2}} \right)^{1/3}, \label{Ana:gas}\\
c_\mathrm{s}=\frac{\pi G \Sigma_\mathrm{gas}Q}{\Omega_{\rm K}}
&=\left(\frac{G\dot{M}_\mathrm{gas}Q}{3\alpha}\right)^{1/3}. \label{Ana:c_s}
\end{align}
An advantage of these expressions is that they do not explicitly involve $T$ and therefore do not explicitly depend on the disk's heating mechanisms. These expressions are useful for self-gravitating disks, where the accretion driven by the GI tends to regulate the value of $Q$ and $\alpha(Q)$ to $\approx 1$--$1.4$ and $\sim 0.1$, respectively (\citealt{Zhu+10}, see also Section~\ref{sec:gas}).

When collisional fragmentation limits dust growth, the grains grow until the collision velocity $\Delta v$ reaches the fragmentation velocity $v_{\rm frag}$ \citep{Birnstiel+11,Birnstiel+2012}. Therefore, the fragmentation-limited grain radius $a_{\rm frag}$ is determined by the relation
\begin{equation}\label{ana:frag}
\Delta v(a_{\rm frag}) = v_\mathrm{frag}.
\end{equation}
Using Equations \eqref{deltav}, (\ref{Ana:gas}), and (\ref{Ana:c_s}),
one can derive the fragmentation-limited radius as
\begin{subequations}\label{ana:a_dust}
\begin{equation}
a_\mathrm{frag} =\max\left( a_{\rm frag, 1}, a_{\rm frag, 2} \right),\label{ana:a_dust0}
\end{equation}
where
\begin{equation}\label{ana:a_dust1}
\begin{split}
a_{\rm frag, 1}
&= \frac{v_{\rm frag}}{\pi \rho_{\rm int} c_{\rm s}} \left( \frac{512 m_{\rm gas} \Sigma_{\rm gas}^{3}}{\sigma_{\rm mol} \alpha^{3}} \right)^{1/4}
\\
&\approx \frac{29}{\alpha^{2/3}Q^{5/6}}
\left(\frac{\rho_\mathrm{int}}{3\:\mathrm{g~cm^{-3}}} \right)^{-1}
\left(\frac{v_\mathrm{frag}}{10\:\mathrm{m\:s^{-1}}} \right)\\
&\times \left(\frac{\dot{M}_\mathrm{gas}}{10^{-4}\:M_\odot\:\mathrm{yr}^{-1}} \right)^{-1/12}
\left(\frac{M}{10\:M_\odot} \right)^{3/8}
\left(\frac{r}{100\:\mathrm{au}} \right)^{-9/8} \micron,
\end{split}
\end{equation}
\begin{equation}\label{ana:a_dust2}
\begin{split}
a_{\rm frag, 2} &= \frac{v_\mathrm{frag}^2\Sigma_\mathrm{gas}}{1.15\pi \rho_\mathrm{int} \alpha c_\mathrm{s}^2}\\
&\approx
\frac{53}{\alpha^{2/3}Q^{4/3}}
\left( \frac{\rho_\mathrm{int}}{3\:\mathrm{g~cm^{-3}}} \right)^{-1}
\left( \frac{v_\mathrm{frag}}{10\:\mathrm{m\:s^{-1}}} \right)^{2}\\
&\times \left(\frac{\dot{M}_\mathrm{gas}}{10^{-4}\:M_\odot\:\mathrm{yr}^{-1}} \right)^{-1/3}
\left( \frac{M}{10\:M_\odot} \right)^{1/2}
\left(\frac{r}{100\:\mathrm{au}} \right)^{-3/2}
\micron,
\end{split}
\end{equation}
\end{subequations}
are the expressions for $a_{\rm frag}$ for ${\rm St}\ll{\rm Re}^{-1/2}$ and $\gg{\rm Re}^{-1/2}$ ($\Delta v = \Delta v_1$ and $\Delta v_2$), respectively.
For self-gravitating disks with $Q =1$--$1.4$, the factors $\alpha^{-2/3}Q^{-5/6}$ and $\alpha^{-2/3}Q^{-4/3}$ in Equations (\ref{ana:a_dust1}) and (\ref{ana:a_dust2}) fall in the range of $\sim2$--$8$.

The corresponding Stokes numbers are
\begin{subequations}\label{Ana:St}
\begin{equation}\label{Ana:St0}
{\rm St_{frag}}=\max\left( {\rm St_{frag,1}}, {\rm St_{frag,2}}\right),
\end{equation}
where
\begin{equation}
\begin{split}
{\rm St_{frag,1}}&=\frac{\pi \rho_{\rm int} a_{\rm frag, 1}}{2\Sigma_{\rm gas}}\\
&\approx
\frac{8.6\times10^{-5}}{\alpha^{1/3} Q^{3/2}} \left(\frac{v_\mathrm{frag}}{10\:\mathrm{m\:s^{-1}}}\right)\\
&\times \left(\frac{\dot{M}_\mathrm{gas}}{10^{-4}\:M_\odot\:\mathrm{yr}^{-1}}\right)^{-5/12}
\left(\frac{M}{10\:M_\odot} \right)^{-1/8}
\left(\frac{r}{100\:\mathrm{au}} \right)^{3/8},
\label{ana:St1}
\end{split}
\end{equation}
\begin{equation}
\begin{split}
{\rm St_{frag,2}}&=\frac{\pi \rho_{\rm int} a_{\rm frag, 2}}{2\Sigma_{\rm gas}}\\
&\approx
\frac{1.6\times10^{-4}}{\alpha^{1/3} Q^{1/6}}
\left(\frac{v_\mathrm{frag}}{10\:\mathrm{m\:s^{-1}}}\right)^2
\left(\frac{\dot{M}_\mathrm{gas}}{10^{-4}\:M_\odot\:\mathrm{yr}^{-1}}\right)^{-2/3}.
\label{ana:St2}
\end{split}
\end{equation}
\end{subequations}

Because ${\rm St}$ determines $v_{r, \rm dust}$, one can also derive estimate $\Sigma_{\rm dust} = \dot{M}_{\rm dust}/(2\pi r|v_{r, \rm dust}|)$.
As shown below, the dust-to-gas surface density ratio $\Sigma_{\rm dust}/\Sigma_{\rm gas}$ in  massive protostellar disks tends to become radially uniform because $v_{r, \rm dust} \approx v_{r, \rm gas}$. 
From Equations (\ref{eq:vgr}) and (\ref{v_r_dust}) with ${\rm St}\ll1$,
the relative difference in the dust and gas radial velocities can be described as
\begin{equation}\label{eq:raddrift}
\left|\frac{v_{r, \mathrm{dust}} -  v_{r, \mathrm{gas}}}{v_{r,\rm gas}}\right|
=\frac{2}{3}\left|\frac{d\mathrm{ln}P}{d\mathrm{ln}r}\right| \frac{\mathrm{St}}{\alpha} \sim \frac{{\rm St}}{\alpha}
\end{equation}
For typical parameters for massive protostellar disks, Equation \eqref{ana:St1} and \eqref{ana:St2} show that ${\rm St}_{\rm frag} \ll \alpha_{\rm floor} < \alpha$, indicating that the radial velocity difference between the gas and dust is small.
Dust vertical settling is also negligible (Equation (\ref{eq:Hdust})).

\section{Numerical Results} \label{sec:Result}
Here we present the results of the numerical calculations for the models listed in Table~\ref{tab:model}.
For all models presented here, the entire part of the disk has temperatures above 200 K and thus can be regarded as being inside the water snowline.

\subsection{The Fiducial Model}\label{subRe:fd}

\begin{figure*}
\gridline{\fig{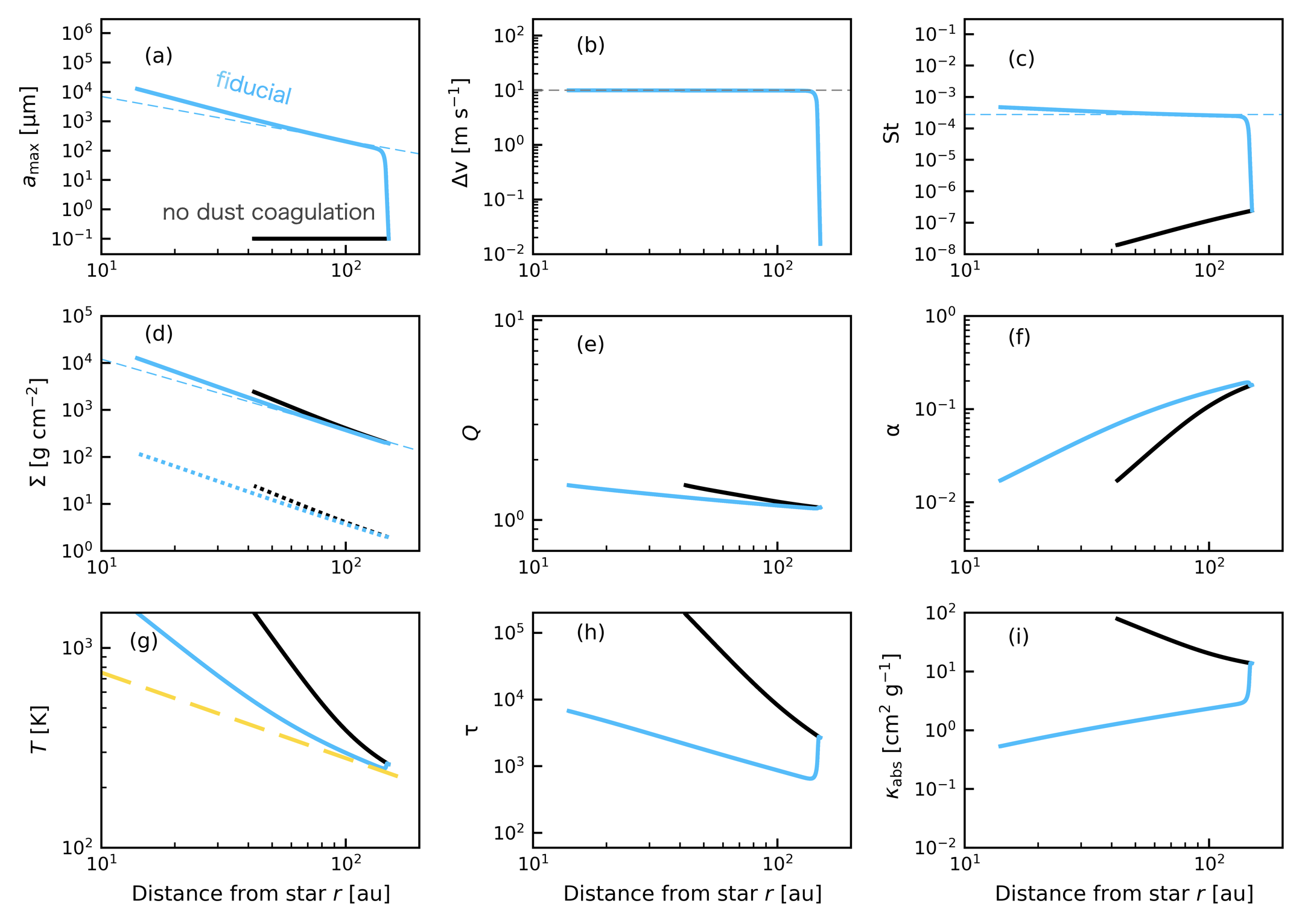}{1\textwidth}{}}
\caption{
Radial profiles of physical quantities in the fiducial model (thick light-blue lines):
(a) maximum dust radius $a_{\rm max}$,
(b) dust collisional velocity $\Delta v$,
(c) Stokes number $\rm St$,
(d) surface densities of gas $\Sigma_\mathrm{gas}$ (solid) and dust $\Sigma_{\rm dust}$ (dashed),
(e) Toomre parameter $\mathrm{Q}$,
(f) viscous parameter $\alpha$,
(g) temperature $T$,
(h) vertical optical depth $\tau$,
and (i) dust opacity $\kappa_\mathrm{abs}$.
For comparison, the profiles without dust evolution (i.e., $a_{\rm min}=a_{\rm max}=0.1\:\micron$) are shown in solid black lines in all panels, except for panel (b).
In panels (a), (c), and (d), the thin dashed light-blue lines represent the analytic solutions, assuming $Q=1.2$ as a typical value (see Section \ref{sec:Analytic}).
In panel (b), the thin gray dashed line shows the assumed fragmentation velocity of $10{\rm\:m\:s^{-1}}$ for reference.
In panel (g), the yellow dashed line represents the irradiation-dominated temperature $T_{\rm irr}$ for reference.
\label{fig:fiducial1}}
\end{figure*}

Figure\:\ref{fig:fiducial1} shows the steady-state disk structure for the fiducial disk model.
Figure\:\ref{fig:fiducial1}(a)--(c) plot the maximum grain size $a_{\mathrm{max}}$, collision velocity $\Delta v$, and the Stokes number $\rm St$ as a function of radial distance $r$. These plots indicate that the grain growth in this model is limited by  collisional fragmentation.
At the disk's outer edge lying at 150 au, the dust grains supplied from the envelope grow locally to $\sim100{\rm\:\micron}$ in size (Figure\:\ref{fig:fiducial1}(a)).
This local growth terminates as the collisional velocity $\Delta v$ reaches the fragmentation velocity $v_{\rm frag}$  (Figure\:\ref{fig:fiducial1}(b)).
After that, the grains accrete inward, keeping the size determined by the balance between coagulation and fragmentation ($\Delta v\approx v_{\rm frag}$).
The Stokes number of the inward accreting grains is $\sim 10^{-4}$, nearly independent of $r$ (Figure\:\ref{fig:fiducial1}(c)). 
This value of $\mathrm{St}$ is much smaller than $\alpha_{\rm floor} =10^{-2}$, indicating that the grains' radial velocity is dominated by the gas accretion velocity, as mentioned in Section~\ref{sec:Analytic}.

The disk structure is regulated by the GI torque.
Figure\:\ref{fig:fiducial1}(d)--(f) present the gas and dust surface densities, Toomre parameter $Q$, and viscous parameter $\alpha$ in the fiducial model.
Both the gas and dust surface densities increase toward smaller $r$, roughly following a single power of $r^{-1.5}$ (Figure\:\ref{fig:fiducial1}(d)).
The radial dependence of $\Sigma_{\rm gas}$ agrees with our analytic estimate, Equation\:(\ref{Ana:gas}), showing $\Sigma_{\rm gas} \propto \Omega_{\rm K} \propto r^{-3/2}$.
The entire disk is marginally unstable at $Q\sim1$,
inducing a GI torque of $\alpha\sim0.02$--$0.2$ (Figure\:\ref{fig:fiducial1}(e) and (f)).
Because $\alpha \gg \rm St$,
the gas and dust accrete toward the central star at nearly the same velocity  (see Section~\ref{sec:Analytic}), resulting in a radially constant dust-to-gas surface density ratio of $\Sigma_{\rm dust
}/\Sigma_{\rm gas} \approx f_\mathrm{ dg, ISM}=0.01$ (Figure\:\ref{fig:fiducial1}(d)).

Because $Q$ is radially nearly constant, the analytic estimates derived in Section~\ref{sec:Analytic} can be used to predict $a_{\rm max}$ and $\rm St$ for the accreting grains.
This is demonstrated in Figure\:\ref{fig:fiducial1}(a) and (c), where the dashed lines show Equations (\ref{ana:a_dust}) and (\ref{Ana:St}) for $Q=1.2$.
We find that the analytic estimates well reproduce the numerical results at $r\ga 40~\rm au$. 
The agreement is less good at $r \la 40~\rm au$ because the actual value of $Q$  depends weakly on $r$. 
Since we are primarily interested in the outer $\sim100{\rm\:au}$ region, which is the main target of radio imaging observations, we use $Q=1.2$ as the reference value for self-gravitating disks.

Figure\:\ref{fig:fiducial1}(g)--(i) show the temperature $T$, optical depth $\tau$, and opacity $\kappa_{\rm abs}$ for the fiducial model.
Around $r\sim 100{\rm au}$, the disk heating is dominated by stellar irradiation, yielding $T \approx T_{\rm irr}$ (Figure\:\ref{fig:fiducial1}(g)).
The temperature appreciably exceeds  $T_{\rm irr}$ at $r \ll 100~\rm au$, where the accretion heating dominates owing to the deeper gravitational potential and higher optical depth (Figure\:\ref{fig:fiducial1}(h)).
Because the maximum grain size $a_{\rm max}$ increases as the grains accrete toward the central star, the opacity decreases toward smaller $r$ (Figure\:\ref{fig:fiducial1}(i)).

To see how dust coagulation contributes to  the disk structure shown above,
we also show in Figure\:\ref{fig:fiducial1} the calculation results for a uniform, constant grain radius of $0.1{\rm\:\micron}$ (see the black lines).
One can see that dust coagulation leads to smaller $\kappa_{\rm abs}$ and $\tau$, resulting in lower $T$ (see Figure\:\ref{fig:fiducial1}(i), (h), (g), respectively).
In contrast, the surface densities $\Sigma_{\rm gas}$ and $\Sigma_{\rm dust}$ are almost unchanged (Figure\:\ref{fig:fiducial1}(d)).
This is consistent with Equation (\ref{Ana:gas}), which shows that  in a self-gravitating disk whose accretion is  regulated by the GI torque ($Q\sim1$ and $\alpha\sim0.1$), the surface density is determined independently of the disk temperature.

\subsection{Dependence on the Fragmentation Velocity}\label{subRe:vfrag}

\begin{figure*}
\gridline{\fig{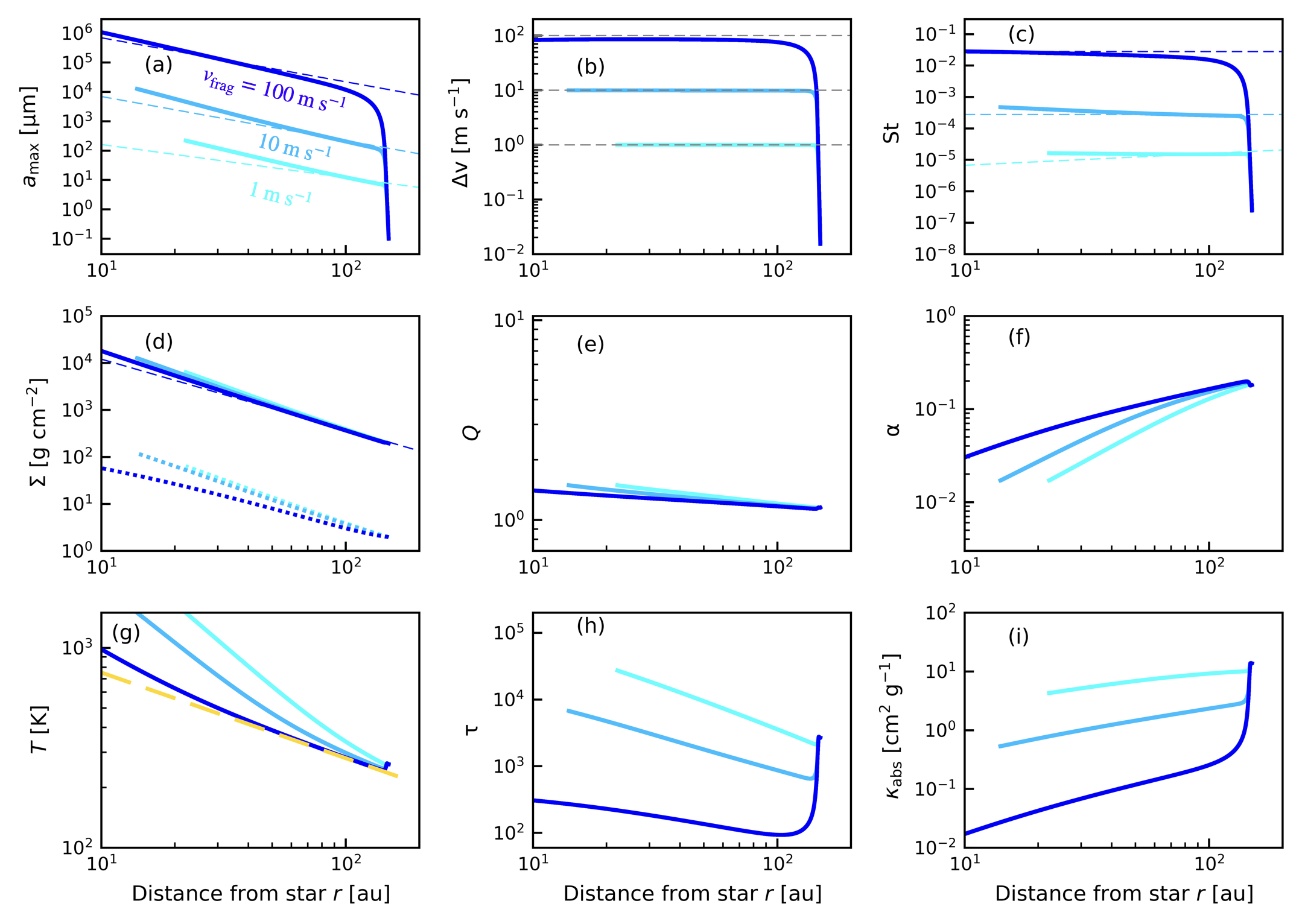}{1\textwidth}{}}
\caption{
Same as Figure \ref{fig:fiducial1},
but for the fiducial and comparison models with $v_{\rm frag}=1$, $10$, and $100{\rm\:m\:s^{-1}}$ (as indicated in panel a).
\label{fig:vfrag}}
\end{figure*}

We here investigate how dust coagulation and the disk structure depend on the fragmentation velocity $v_{\rm frag}$.
Figure\:\ref{fig:vfrag} shows the results for $v_\mathrm{frag}=1$, $10$ (fiducial), and $100{\rm\:m\:s^{-1}}$.

All three cases show $\Delta v\approx v_{\rm frag}$ (Figure\:\ref{fig:vfrag}(b)), suggesting that fragmentation mainly limits dust coagulation.
The radial drift barrier only gives a minor contribution to the coagulation limit, although its effect is appreciable in the case of $v_{\rm frag} = 100~\rm m~s^{-1}$, which shows that $\Delta v$ is slightly lower than $v_{\rm frag}$.
Figure\:\ref{fig:vfrag}(a) and (c) show that a higher $v_{\rm frag}$ leads to a larger maximum grain size and a larger Stokes number,
consistent with the analytic estimates given by Equations (\ref{ana:a_dust}) and (\ref{Ana:St}).

The gas surface density is insensitive to the fragmentation velocity (Figure\:\ref{fig:vfrag}(d)). This is because the disk accretion is self-regulated to the marginally unstable state with $Q\sim1$ and $\alpha\sim0.1$ (Figure\:\ref{fig:vfrag}(e)--(f)), which only implicitly depends on $T$ (see also Equation\:\eqref{Ana:gas}).
The relation $\Sigma_{\rm rm dust} \approx 0.01\Sigma_{\rm gas}$ seen in the fiducial model approximately holds as long as $v_{\rm frag} < 100~\rm m~s^{-1}$. For $v_{\rm frag} = 100~\rm m~s^{-1}$, $\Sigma_{\rm  dust}$ is slightly shallower than $\Sigma_{\rm  gas}$ due to the grains' non-negligible radial inward drift relative to the gas.

The disk temperature is lower with higher $v_{\rm frag}$ (Figure\:\ref{fig:vfrag}(g)),
since the optical depth and opacity are lower with larger grain size (Figure\:\ref{fig:vfrag}(h)--(i)).
In the most optically-thin case of $v_{\rm frag}=100{\rm\:m\:s^{-1}}$, stellar irradiation is the dominant heat source at $r\ga30{\rm\:au}$.
As mentioned above,
any change in temperature has little impact on the gas surface density, which is the characteristic of self-gravitating disks.

\subsection{Dependence on the Accretion Rate}\label{subRe:Mdot}

\begin{figure*}
\gridline{\fig{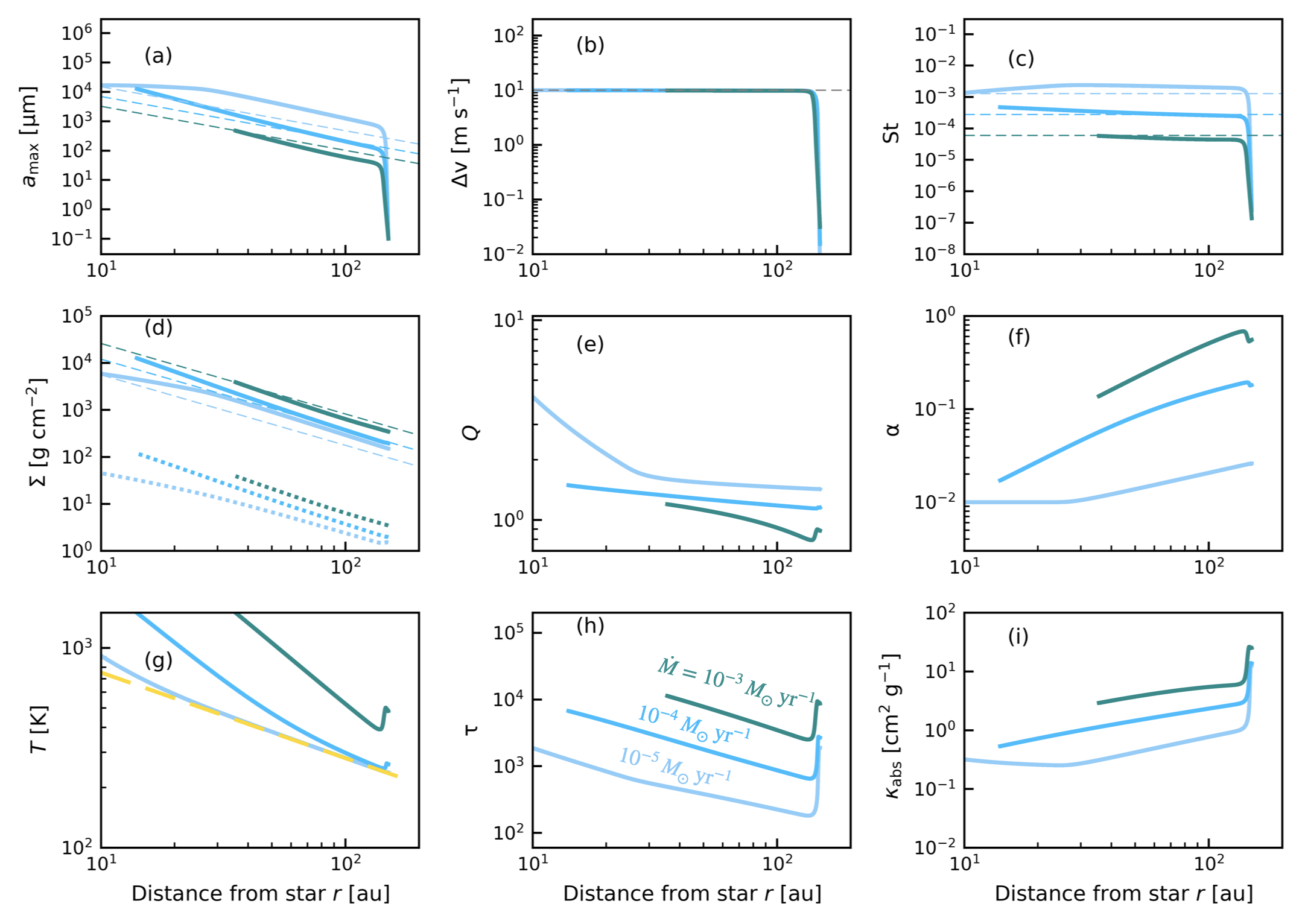}{1\textwidth}{}}
\caption{
Same as Figure \ref{fig:fiducial1},
but for the fiducial and comparison models with $\dot{M}_{\rm gas}=10^{-5}$, $10^{-4}$, and $10^{-3}\:M_\odot{\rm\:yr^{-1}}$ (as indicated in panel h).
\label{fig:Mdot}}
\end{figure*}

Next, we present the model calculations with various accretion rates, fixing the  fragmentation velocity to $v_{\rm frag}=10{\rm\:m\:s^{-1}}$.
Figure\:\ref{fig:Mdot} shows the results for $\dot{M}_{\rm gas}=10^{-5}$, $10^{-4}$ (fiducial), and $10^{-3}\:M_\odot{\rm\:yr^{-1}}$.

In all models, dust coagulation is regulated by the collisional fragmentation  (Figure\:\ref{fig:Mdot}(b)).
The maximum grain size is smaller for higher accretion rates (Figure\:\ref{fig:Mdot}(a)).
The main reason for this trend is that
a higher $\dot{M}_{\rm gas}$ leads to higher $\alpha$ and $T$  (see Figure\:\ref{fig:Mdot}(f)--(g)), both of which increase the turbulence-induced collision velocity $\propto \alpha^{1/2} c_{\rm s} \propto (\alpha T)^{1/2}$.
However, the $\dot{M}_{\rm gas}$ dependence of the maximum grain size is minor
compared with the $v_{\rm frag}$ dependence shown in Figure\:\ref{fig:vfrag}.

The disk with higher $\dot{M}_{\rm gas}$ has higher surface-densities $\Sigma_{\rm gas}$ and $\Sigma_{\rm dust}$ (Figure\:\ref{fig:Mdot}(g)).
However, the dependence is rather weak, with a factor of 100 increase in $\dot{M}_{\rm gas}$ only resulting in a factor of several increases in $\Sigma_{\rm gas}$ and $\Sigma_{\rm dust}$.
Because $a_{\rm max}$ increases with $\dot{M}_{\rm gas}$, a higher $\dot{M}_{\rm gas}$ yields a lower $\kappa_{\rm abs}$ (Figure\:\ref{fig:Mdot}(i)). The increase in $\kappa_{\rm gas}$ and $\Sigma_{\rm gas}$ results in larger $\tau$ and hence higher $T$ for higher $\dot{M}_{\rm gas}$ (Figure\:\ref{fig:Mdot}(g)--(h)).

In the case of $\dot{M}_{\rm gas} = 10^{-5}\:M_\odot{\rm\:yr^{-1}}$, the regions inside and outside $r \approx 30{\rm\:au}$ exhibit distinct radial structures.
This is because, in this low-$\dot{M}_{\rm gas}$ disk, the inner part is gravitationally stable with $Q>1.4$ (Figure\:\ref{fig:Mdot}(e))
and $\alpha \approx \alpha_{\rm floor}=0.01$ (Figure\:\ref{fig:Mdot}(f)).

In the lower and higher $\dot{M}$ models, the numerical results deviate more from the analytic estimates than in the fiducial model,
because the $Q$ values are greater or less than the reference value of $1.2$ in those cases (Figure\:\ref{fig:Mdot}(e)).
However, the deviations from the analytic estimates are only a factor of a few at the outer region of $\sim100{\rm\:au}$.

\section{Discussion} \label{sec:discc}
\subsection{Constraining the Fragmentation Velocity from Observations of Massive Protostellar Disks}\label{giron:1}
In the previous section, we have shown that collisional fragmentation limits the growth of silicate grains in massive protostellar disks.
The maximum grain size of the silicate grains is primarily determined by their threshold fragmentation velocity $v_{\rm frag}$. The maximum size is only weakly dependent on the accretion rate and, more importantly,  independent of the disk temperature owing to the self-regulating nature of the self-gravitating disks (see also Section~\ref{sec:Analytic}). 
These suggest that measurements of the maximum grain size in massive protostellar disks can be used to constrain $v_{\rm frag}$ for silicates, which is currently highly uncertain, as introduced in Section \ref{sec:intro}.
In this subsection, we derive analytic formulas that are useful for this purpose. An application of the formulas is presented in Section~\ref{sec:GGD27}. 

Equations (\ref{ana:a_dust0})--(\ref{ana:a_dust2}) in Section~\ref{sec:Analytic} provide analytic estimates for the fragmentation-limited grain size $a_{\rm max}$ for given $v_{\rm frag}$.
Now we invert them into formulas that return $v_{\rm frag}$ for given $a_{\rm max}$. The result reads
\begin{subequations}\label{dis:vfrag}
\begin{align}
v_\mathrm{frag} = &\min \left( v_\mathrm{frag,1}, v_\mathrm{frag,2} \right),\\
\begin{split}
v_\mathrm{frag,1} &\approx
35\alpha^{2/3} Q^{5/6}\left(\frac{a_\mathrm{max}}{100\:\rm{\mu m}} \right)
\left(\frac{\rho_\mathrm{int}}{3\:\mathrm{g~cm^{-3}}} \right) \\
&\times \left(\frac{\dot{M}_\mathrm{gas}}{10^{-4}\:M_\odot\:\mathrm{yr}^{-1}} \right)^{1/12}
\left(\frac{M}{10\:M_\odot} \right)^{-3/8}
\left(\frac{r}{100\:\mathrm{au}} \right)^{9/8}\:\mathrm{m\:s^{-1}}, \label{dis:vfrag1}
\end{split}
\\
\begin{split}
v_\mathrm{frag,2} &\approx
14\alpha^{1/3} Q^{2/3}\left(\frac{a_\mathrm{max}}{100\:\mathrm{\mu m}} \right)^{1/2}
\left(\frac{\rho_\mathrm{int}}{3\:\mathrm{g~cm^{-3}}} \right)^{1/2} \\
&\times \left(\frac{\dot{M}_\mathrm{gas}}{10^{-4}\:M_\odot\:\mathrm{yr}^{-1}} \right)^{1/6}
\left(\frac{M}{10\:M_\odot} \right)^{-1/4}
\left(\frac{r}{100\:\mathrm{au}} \right)^{3/4}\:\mathrm{m\:s^{-1}}.\label{dis:vfrag2}\end{split}
\end{align}
\end{subequations}
Putting $Q=1.2$ and $\alpha(Q=1.2)=0.14$ for typical self-gravitating disks and $\rho_{\rm int}=3.0{\rm\:g\:cm^{-3}}$ for silicates, we have  
\begin{subequations}\label{ana:vfrag}
\begin{align}
&v_\mathrm{frag} = \min\left( v_{\rm frag,1}, v_{\rm frag,2} \right),\\ 
\begin{split}
&~~~v_\mathrm{frag,1} \approx 
 11  \left(\frac{a_\mathrm{max}}{100\:\rm{\mu m}} \right)
\left(\frac{M}{10\:M_\odot} \right)^{-3/8} \\
&~~~~~~~~~~~~~~~~\times
\left(\frac{\dot{M}_\mathrm{gas}}{10^{-4}\:M_\odot\:\mathrm{yr}^{-1}}\right)^{1/12}
\left(\frac{r}{100\:\mathrm{au}} \right)^{9/8}\:\mathrm{m\:s^{-1}},
\end{split}
\\
\begin{split}
&~~~v_\mathrm{frag,2} \approx
7.9 \left(\frac{a_\mathrm{max}}{100\:\rm{\mu m}} \right)^{1/2}
\left(\frac{M}{10\:M_\odot} \right)^{-1/4} \\
&~~~~~~~~~~~~~~~~\times
\left(\frac{\dot{M}_\mathrm{gas}}{10^{-4}\:M_\odot\:\mathrm{yr}^{-1}}\right)^{1/6}
\left(\frac{r}{100\:\mathrm{au}} \right)^{3/4}\:\mathrm{m\:s^{-1}}.
\end{split}
\end{align}
\end{subequations}
Importantly, these estimates depend only weakly on the gas accretion rate and stellar mass, which are difficult to determine precisely from observations.

\subsection{The Case of the GGD27-MM1 Disk and Its Implication for Rocky Planetesimal Formation} \label{sec:GGD27}

\begin{figure}[t!]
\begin{center}
\includegraphics[width=8.5cm, bb=0 0 1079 1080]{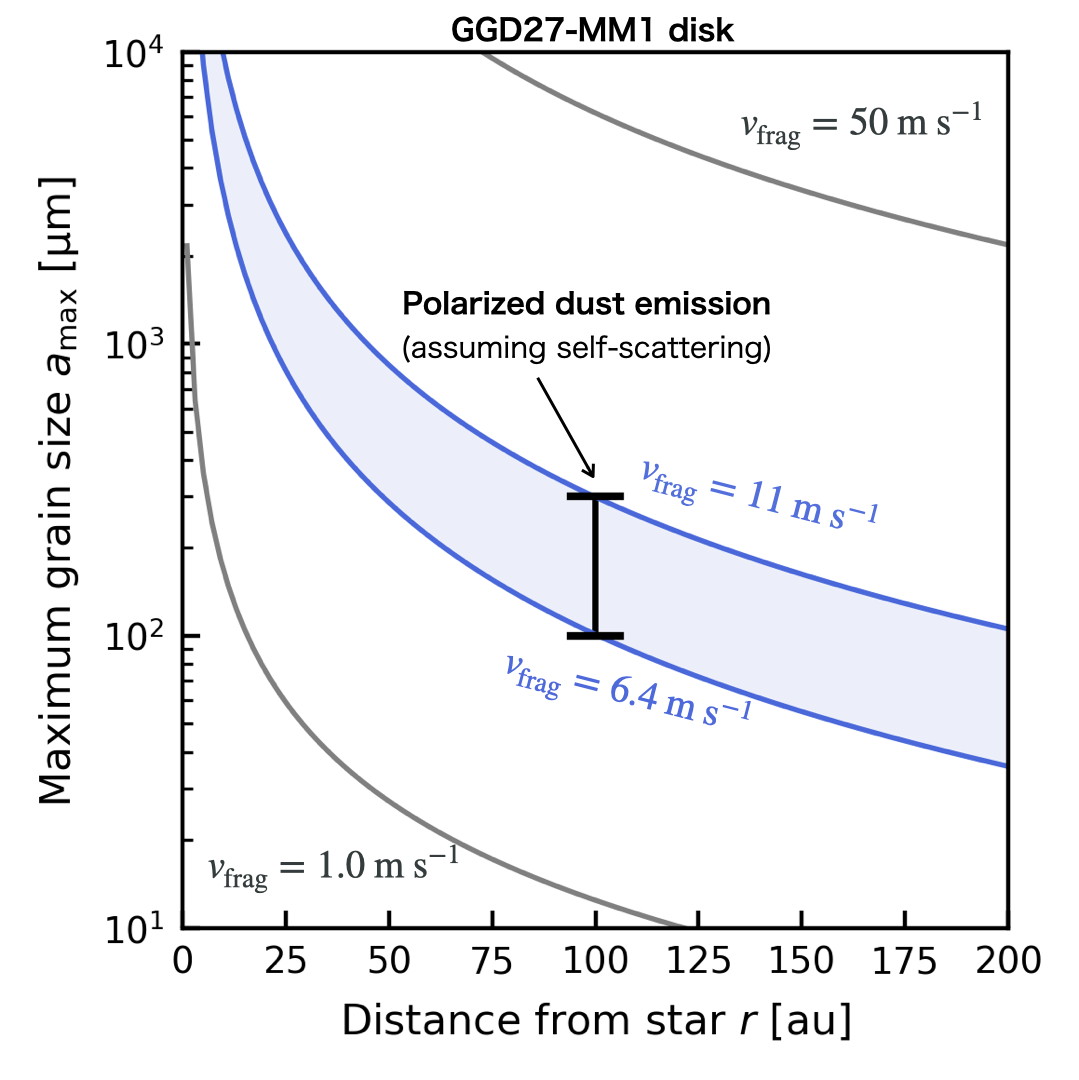}
\end{center}
\caption{
Fragmentation-limited grain radii $a_{\rm frag}$ (Equation (\ref{ana:frag})) for the GGD27-MM1 disk (see text for the adopted stellar parameters) with different values of the fragmentation threshold $v_{\rm frag}$. The vertical bar indicates the inferred range of the maximum grain radius $a_{\rm max}$ at $r = 100~\rm au$ from the assumption that the polarized emission from this region is due to dust self-scattering. The inferred $a_{\rm max}$ is consistent with $v_{\rm frag}=6.4$--$11{\rm\:m\:s^{-1}}$ (blue shaded area) and rules out $v_{\rm frag}=1$ and $50{\rm\:m\:s^{-1}}$ (gray lines).
\label{fig:Giron:ana}} 
\end{figure}

To date, the disk of GGD27-MM1 is the only massive protostellar disk for which a measurement of the maximum grain size from observations is available  \citep{Girart+18}.
We here apply Equation (\ref{ana:vfrag}) to the observations to infer the fragmentation velocity for the silicates in this disk.

The 1.14 mm polarimetric observations of the GGD27-MM1 disk \citep[Figure 3 of][]{Girart+18} showed that 
the disk emission around  $r \approx 100{\rm\:au}$ is polarized, with  the polarization vector aligned with the disk's minor axis.
Such polarization features can be explained by the self-scattering of the thermal emission by dust grains with a maximum radius of $\sim 100{\:\micron}$ \citep{Kataoka+16,Yang+2017}.
In principle, the grain size constraint derived from self-scattering models may not reflect the size of the grains lying at the disk midplane if the disk is optically thick and if larger grains have settled to the midplane \citep{UedaKataoka21}. 
However, as discussed in Section~\ref{sec:Analytic},  dust settling in massive protostellar disks should be negligible (${\rm St} \ll \alpha$), and thus the grains responsible for the polarized emission should also represent the grains at the midplane.

To give a more quantitative estimate of the maximum grain size $a_{\rm max}$, we focus on the polarized emission at $r = 100\rm\:au$ in the GGD27-MM1 disk. According to Figure 3 of \citet{Girart+18}, the emission from this position has a polarization degree of $\sim 1.5\%$
(we here assume the source distance of 1.4 kpc; \citealt{Anez+2020}).
We then follow \citet{Girart+18} and convert the polarization degree into $a_{\rm max}$ based on the self-scattering model of \citet[][their Figure 3]{Kataoka+16}. We estimate $a_{\rm max} \approx 100$--$300\:\micron$ for the polarization degree of $\sim 1.5\%$. 
Our estimate here is still crude because the self-scattering model we use here was not originally designed to predict the polarization degree of the GGD27-MM1 disk. 
Moreover, the model of \citet{Kataoka+16} assumes perfectly spherical grains, but more recent studies show that the polarization degree also depends on the shape of the grains \citep{Kirchschlarger22,Tazaki22,Lin+2023}. 
Future dedicated radiative transfer modeling of this particular system will allow a more accurate estimate for $a_{\rm max}$.

We now use Equation (\ref{ana:vfrag}) to infer the fragmentation velocity of silicates in the GGD27-MM1 disk. 
We assume $M=20\:M_\odot$ and $\dot{M}_{\rm gas} = 7\times10^{-5}\:M_\odot{\rm\:yr}^{-1}$ for GGD27-MM1 \citep{Anez+2020}.
Equation (\ref{ana:vfrag}) then yields $v_{\rm frag}\approx 6.4$--$11{\rm\:m\:s^{-1}}$ for the estimated maximum grain size of $a_{\rm amax} \approx 100$--300 $\micron$ at $r=100{\rm\:au}$.
This is also shown in Figure \ref{fig:Giron:ana}, which plots the fragmentation-limited grain size $a_{\rm frag}$ (Equation (\ref{ana:a_dust})) for various $v_{\rm frag}$ values.
Importantly, our estimate rules out the commonly quoted value of $v_{\rm frag} = 1~\rm m~s^{-1}$  for silicates, along with the most sticky scenario of $v_{\rm frag} = 50~\rm m~s^{-1}$ in the literature (see Section~\ref{sec:intro}).
It is worth noting that our work is consistent with the conclusion drawn by \cite{Liu2021}, who similarly estimated a fragmentation velocity of $\sim 10\: \rm{m\:s^{-1}}$ for silicate dust based on observations of the accretion burst FU Ori Disk.
For completeness, we show in Appendix~\ref{sec:appA} the radial structure of the GGD27-MM1 disk obtained from direct numerical calculations of the model equations, confirming that $v_{\rm frag}\approx 10{\rm\:m\:s^{-1}}$ best explains the polarized emission at $r = 100~\rm au$. 

The above exercise is only one example based on the observations of a single system at a single wavelength. Clearly, more case studies with other massive protostellar disks and at various wavelengths are needed to give a more robust constraint on $v_{\rm frag}$ for silicates.
With this caveat in mind,
it is still interesting to discuss what the derived fragmentation velocity of $v_{\rm frag}\approx 10{\rm\:m\:s^{-1}}$ would imply for the formation of rocky planetesimals around low-mass protostars.
As already mentioned in Section~\ref{sec:intro}, the maximum value of grain collisional velocity in protoplanetary disks is $\approx 20$--$50{\rm\:m\:s^{-1}}$, even if no strong turbulence is present \citep{Johansen+2014}.
With $v_{\rm frag}\approx10{\rm\:m\:s^{-1}}$,
aggregates of silicate grains in protoplanetary disks are likely to experience catastrophic fragmentation at some point, meaning that rocky planetesimals would not form solely via the coagulation of silicates.
The streaming and gravitational instabilities would be necessary to account for rocky planetesimal formation.
Interestingly, the fragmentation velocity we derived is still considerably higher than the commonly assumed value of 1 $\rm m~s^{-1}$ \citep{Guttler10}. Because the streaming and gravitational instabilities favor large grains whose aerodynamical coupling to the surrounding gas is moderate, our estimate for $v_{\rm frag}$ suggests that rocky planetesimal formation via these mechanisms may be more efficient than previously anticipated.

\subsection{Caveats of Our Model}
\label{sec:caveats}

This work is only the first step toward a full understanding of dust evolution in massive protostellar disks and rests on several assumptions and simplifications.
Here, we describe some caveats of our model.

Our model uses Epstein's drag law to compute the dust stopping time and thus the Stokes number.
We have confirmed that the condition for Epstein's law, $a_{\rm max} \ll \lambda_\mathrm{mfp}$, holds in all disk models presented in this study, except in the inner region of $r<20{\rm\:au}$ with the high fragmentation velocity of $v_{\rm frag}=100{\rm\:m\:s^{-1}}$.
Therefore, this assumption does not affect our main conclusions. In any case, it is straightforward to extend our model beyond the Epstein drag regime. 

We have focused on the steady part of the disk structure by enforcing steady accretion.
However, self-gravitating disks are often highly time-varying due to accretion bursts \citep[e.g.,][]{Meyer+17}.
\citet{Vorobyov+22} simulated dust collisional evolution during accretion bursts in low-mass protostellar disks.
In their simulations, the snowline moves away from a few au to several dozen of au during bursts due to increased stellar luminosity.
The snowline excursion would induce a drop in the maximum grain size in the temporal silicate-dust region if bare silicate grains are more fragile than ice grains. 
However, in massive protostellar disks, the snowline already lies at as far as several hundred au from the central star even without bursts.
In these disks, accretion bursts would not affect the grain size in the $r\lesssim{100\rm\:au}$ region.

We have neglected fragmentation of self-gravitating disks,
which in fact could occur if $Q<0.6$ \citep[e.g.,][]{Takahashi+16}.
However, in all disk models presented in Section \ref{sec:Result}, the Toomre parameter is larger than $0.6$ in the entire disks, indicating that  disk fragmentation would be negligible.

We adopted the $\alpha$ value as the function of Toomre parameter $Q$ (Equations (\ref{hanpuku4}) and (\ref{aGI})),
and selected $Q=1.2$ and $\alpha(Q=1.2)=0.14$ as the reference values.
Although this function is known to reproduce the qualitative of self-gravitating protostellar disks \citep[e.g.,][]{Takahashi+2013}, the accuracy of the function has not been evaluated so far.
For instance, if the GGD27-MM1 disk is moderately unstable ($Q\approx1$) but highly turbulent ($\alpha \approx 1$),
Equation (\ref{dis:vfrag}) predicts 
$v_{\rm frag} \approx20{\rm\:m\:s^{-1}}$ instead  of $v_{\rm frag} \approx 10{\rm\:m\:s^{-1}}$.
Because this fragmentation velocity is comparable to the low end of the range of the maximum grain collision velocity (20--50 $\rm m~s^{-1}$), planetesimal formation by dust coagulation alone may be barely possible with this higher value of $v_{\rm frag}$.

As for the $\alpha$ parameter, we also note that
it is non-trivial whether turbulence and angular-momentum transfer can be represented by a single value of $\alpha$,
especially in self-gravitating disks.
For completeness, we re-derive the formula of the fragmentation velocity, distinguishing two $\alpha$ values:
$\alpha_{\rm turb}$ represents turbulence intensity
and $\alpha_{\rm acc}$ causes accretion through angular momentum transport,
\begin{equation}\label{eq:vfrag_alpha}
v_\mathrm{frag} =\min\left( v_{\rm frag, 1}\left(\frac{\alpha_\mathrm{turb}}{\alpha_\mathrm{acc}}\right)^{3/4},
~v_{\rm frag, 2} \left(\frac{\alpha_\mathrm{turb}}{\alpha_\mathrm{acc}} \right)^{1/2}\right).
\end{equation}
This is equivalent to Equation (\ref{ana:vfrag}) if $\alpha_{\rm turb}=\alpha_{\rm acc}$.
In self-gravitating disks, non-turbulent (coherent) spiral arms could give a considerable contribution to the gravitational accretion stress. In this case, one would have $\alpha_{\rm turb}<\alpha_{\rm acc}$, and therefore $v_{\rm frag}$ estimated by Equation (\ref{eq:vfrag_alpha}) would be smaller than our reference value of $\approx10{\rm\:m\:s^{-1}}$, 
which would strengthen our conclusion that rocky planetesimal formation via silicate coagulation alone is unlikely (Section \ref{sec:GGD27}).

We evaluate the turbulent Reynolds number $\rm Re$, assuming that molecular viscosity determines the smallest eddy scale.
However, in partially ionized media, the turbulence cutoff scale may be much larger due to magnetohydrodynamic waves and radiative cooling \citep{Xu+2016,Sisbee2021}.
Despite the difficulty in determining this scale, we examine the impact of $\rm Re$ uncertainty.
From Equations\:(\ref{deltav1}) and (\ref{dis:vfrag1}), we find $v_\mathrm{frag,1}\propto{\mathrm{Re}}^{1/4}$.
For example, if $\rm Re$ is four orders of magnitude smaller than the value estimated with molecular viscosity, the fragmentation velocity inferred from the GGD27-MM1 observations would decrease to $\approx1{\rm\:m\:s^{-1}}$, further supporting our conclusion that $v_{\rm frag}$ for rocky grains are not as high as 50 $\rm m~s^{-1}$.

The dust collisional velocity of Equation (\ref{deltav}) is based on the Kolmogorov turbulence model \citep{Ormel+2007}.
However, earlier studies suggested that turbulence energy spectra in astronomical disks may deviate from the Kolmogorov law \citep[e.g.,][]{Iroshnikov1964, Kraichnan1965}.
\citet{Gong+2021} expanded upon this by developing collisional velocity formulas for arbitrary turbulence models represented by power-law spectra.
They found that collisional velocities for the Iroshnikov--Kraichnan turbulence and the turbulence caused by magnetorotational instabilities are higher than those for the standard Kolmogorov turbulence.
If the GGD27-MM1 disk has such a turbulence energy spectrum, we estimate that the fragmentation velocity might be as high as $\sim50$--$150{\rm\:m\:s^{-1}}$.

Finally, we discuss the validity and limitations of the single-size approach and the interpretation of radio polarimetric observations.
In general, the single-size approach well approximates the evolution of the maximum grain size and dust surface density when as long as the grain size distribution is top-heavy, i.e., the largest grains dominate the total dust surface density \citep{Birnstiel+2012, Sato+2016}.
However, the single size approach itself does not derive the size distribution, so we have assumed a power-law size distribution to compute the disk's vertical optical depth (Section~\ref{subsec:op}). 
We expect that this assumption would not affect the results presented in this study, because the temperature of self-gravitating disks is self-regulated such that $Q\sim1$ is maintained.
The adopted size distribution is top-heavy (see footnote \ref{footnote1}), so our model is internally consistent in this respect. 
However, depending on the cascade processes of collisional fragmentation \citep[e.g.,][]{Kobayashi+10},
the grain size distribution may become bimodal rather than single-peaked (H. Kobayashi, priv.~comm.). Our single-size approach would not be adequate for treating such distribution.
The uncertainty in the grain size distribution, as well as in the grain shape, could affect the grain size constraint derived from millimeter polarimetric observations based on the dust self-scattering scenario (see also Section \ref{sec:GGD27}).
Further investigation with detailed treatments of the collisional fragmentation  and radio scattering processes will be required in future work.

\section{Summary} \label{sec:sum}
The threshold fragmentation velocity of silicate grains is a critical parameter to understanding the formation of rocky planetesimals, but it is currently highly uncertain.
Silicate dust regions of $\ga200{\rm\:K}$ are too small to observe in protoplanetary disks around low-mass protostars, while they are observable in disks around hotter massive protostars.
In this study, we for the first time developed a theoretical model of dust evolution in massive protostellar disks, and constrained the fragmentation velocity of silicate dust.

We solved the gas disk structure and the dust coagulation self-consistently.
For the gas disk structure, we took into account both stellar irradiation and accretion heating, together with efficient angular momentum transfer of self-gravitating disks.
For the dust evolution, we considered coagulation and fragmentation, as well as radial drift and vertical settling.
We found that the maximum grain size is limited by the collisional fragmentation rather than by the radial drift in massive protostellar disks.
We also derived the analytic formula of the fragmentation-limited grain size in self-gravitating disks (Equation (\ref{ana:a_dust})), which reproduces the numerical results well.
The maximum grain size is larger for the higher fragmentation velocity, and only weakly depends on the heating mechanism and the accretion rate.
This suggests that measurements of the maximum grain size in massive protostellar disks can be used to constrain the fragmentation velocity.

Based on the results of our analytic and numerical calculations, we derived a new simple formula to constrain the fragmentation velocity, i.e., Equation (\ref{ana:vfrag}).
Using this formula with the maximum size of silicate grains in the massive protostellar disk of GGD27-MM1 \citep{Girart+18},
we estimated the threshold fragmentation velocity of silicate grains as $v_{\rm frag} \approx 10{\rm\:m\:s^{-1}}$.
This obtained fragmentation velocity is lower than the maximum collisional velocity expected in protoplanetary disks around low-mass protostars \citep[e.g.,][]{Johansen+2014}.
This implies that rocky planetesimals form not by dust collisional growth but by other mechanisms, such as the gravitational instability of the dust layer.

We note that measurements of the maximum size of silicate grains in massive protostellar disks are still limited and indefinite.
Further polarization observations at multiple wavelengths toward multiple objects are required to accurately constrain the maximum size of silicate dust, and thus the fragmentation velocity.

\section*{}
The authors are grateful to Hiroshi Kobayashi, Yusuke Tsukamoto, Satoshi Ohashi, and John Bally for useful discussions. This work was supported by Japan Society for the Promotion of Science KAKENHI grant Nos.~JP18H05438, JP19K03926, JP19K03941, JP19K14760, JP20H00182, JP20H01948, JP21H00058, and JP21H01145.


\appendix

\section{Models for the GGD27-MM1 disk} \label{sec:appA}

\begin{figure*}
\gridline{\fig{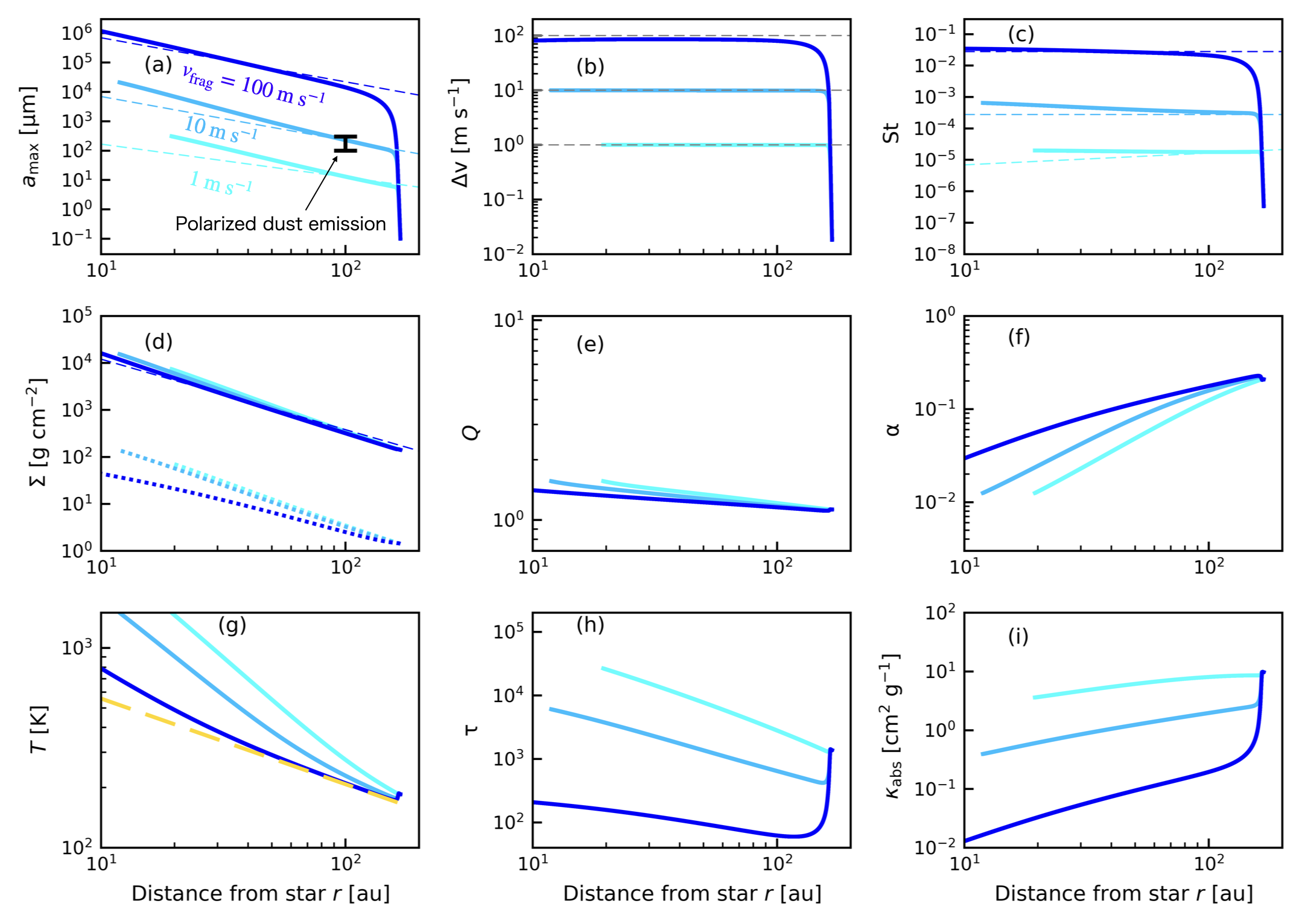}{1\textwidth}{}}
\caption{
Same as Figure\:\ref{fig:fiducial1},
but the parameters are GGD27-MM1(Table\:\ref{tab:model}) and $v_\mathrm{frag}=1,$, $10$, and $100\:\mathrm{m\:s^{-1}}$
(as indicated in panel a).
The thin dashed lines represent the analytic solutions, assuming $Q=1.2$.
For reference, the range of the maximum grain radius suggested by the 1.14-mm polarization observation \citep{Girart+18} is shown by the black vertical line in panel (a).
The model with $v_\mathrm{frag}=10{\rm\:m\:s^{-1}}$ reproduces the observed grain size,
as predicted by the analytic method (see Section \ref{sec:GGD27}).
\label{fig:GGD}}
\end{figure*}

GGD27-MM1 is the massive protostellar disk driving the highly-collimated radio jet HH 80/81.
The previous distance estimate was $1.7{\rm\:kpc}$, but the recent {\it Gaia} DR2 parallax and reddening measurement indicated $1.4{\rm\:kpc}$ \citep{Anez+2020, Zucker+20}.
The several radio knots aligned the jet axis for $\sim10{\rm\:pc}$, and the knot proper motions are as fast as $\sim500$--$1000{\rm\:km\:s^{-1}}$ \citep[e.g.,][]{Marti+95, Masque+15}.
The presence of this outstanding jet makes that GGD27-MM1 is an ideal massive protostellar disk to study in detail.
Moreover, the GGD27-MM1 disk is currently the only protostellar disk for which silicate grain coagulation has been confirmed \citep{Girart+18}.

In Section \ref{sec:GGD27}, we used our simple analytic formula of Equation (\ref{ana:vfrag})
and found that the fragmentation velocity of $v_{\rm frag} \approx 10{\rm\:m\:s^{-1}}$
can reproduce the observed grain size, i.e., $a_{\rm max}=100$--$300{\rm\:\micron}$ at $r=100{\rm\:au}$.
To confirm the evaluated $v_{\rm frag}$ value,
we here perform the detailed numerical model calculations with the parameters of the GGD27-MM1 disk (Table \ref{tab:model}).

Figure \ref{fig:GGD} shows the results of the GGD27-MM1 model calculations with $v_{\rm frag}=1$, $10$, and $100{\rm\:m\:s^{-1}}$.
The basic behaviors of all physical quantities are the same as in the fiducial and comparison models in Section \ref{subRe:vfrag} (see Figure \ref{fig:vfrag}).
The maximum grain size is larger for higher fragmentation velocity.
It can be seen that the model with $v_{\rm frag}=10{\rm\:m\:s^{-1}}$ falls within the range of the observed grain size (panel a),
which confirms our analytic estimation.

\bibliography{ms}{}
\bibliographystyle{aasjournal}

\end{document}